\documentclass[aps,onecolumn]{revtex4}

\usepackage[latin1]{inputenc}
\usepackage[T1]{fontenc}

\usepackage{amsmath}
\usepackage{amsfonts}
\usepackage{amssymb}
\usepackage{array}

\usepackage{multirow}
\usepackage{eurosym}

\usepackage{pslatex}
\usepackage{graphicx}
\usepackage{color}
\usepackage{dcolumn}

\begin{document}

\title{\Large{ELISA: a cryocooled  10 GHz oscillator\\
with $10^{-15}$ frequency stability.}\rm}

\author{S. Grop $^{1}$, P.Y. Bourgeois $^{1}$, N. Bazin $^{1}$, Y. Kersal\'e $^{1}$, E. Rubiola $^{1}$, C. Langham$^{2}$,\\ M. Oxborrow $^{2}$, D. Clapton$^{3}$, S. Walker$^{3}$, J. De Vicente $^{4}$, V. Giordano $^{1}$}

\affiliation{~\\ $^{1}$: FEMTO--ST Institute, \\
Time and Frequency Dpt., UMR 6174 CNRS-ENSMM\\
32 av.~de l'Observatoire, 25044 Besan\c{c}on Cedex, France\\
\vskip 0.5pt
$^{2}$: National Physical Laboratory\\
Queens Road\\
Teddington, Middlesex, TW11 0LW, UK\\
\vskip 0.5pt
$^{3}$: Oxford Instruments plc \\
Tubney Woods\\
Abingdon, Oxon OX13 5QX, UK \\
\vskip 0.5pt
$^{4}$: European Space Agency ESA-ESOC\\
Robert Bosch Str. 5\\
D-64293 Darmstadt, Germany\\}

\vskip 20pt

\begin{abstract}
This article reports the design, the breadboarding and the validation of an
ultra-stable Cryogenic Sapphire Oscillator operated in an autonomous cryocooler.
The objective of this project was to demonstrate the feasibility of a frequency
stability of~ $3\times 10^{-15}$ between 1 s and 1,000 s for the European Space
Agency deep space stations. This represents the lowest fractional frequency
instability ever achieved with cryocoolers.
The preliminary results presented in this paper validate the design we adopted for the sapphire resonator, the cold source and the oscillator loop. 
\end{abstract}

\maketitle

\clearpage
\section{INTRODUCTION}

The ever increasing need for better tracking data and scientific return in deep
space missions calls for the development of new frequency references of improved
stability. The ground station of the European Space Agency (ESA) are currently
equipped with hydrogen masers (HM) which are the most stable commercial atomic clocks
around 1000 s -- 1 day timescales.
Cryogenic Sapphire Oscillators (CSO) offer unbeatable stability performances in timescales ranging from milliseconds to a few hundred seconds, and therefore extremely low phase noise close to the carrier. A combined CSO and HM system in ESA deep space stations would allow to benefit from excellent short term and long stabilities. Compared to the current performance available from hydrogen masers and state of the art quartz oscillators, CSO would provide the means to improve the orbit determination and would open the field to new radio science experiments.
The objective of the project ELISA funded by ESA is the design, the
breadboarding and the validation of a CSO for ground station use, in line with
a frequency stability specification of $\sigma_{y}(\tau) \leq 3\times 10^{-15}$
for $1\mathrm{s}\leq \tau \leq 1,000$s associated with a large autonomy,
{\it{i.e.}} at least one year of continuous operation without maintenance. \\

Such a frequency performance has been already demonstrated by few laboratories operating Whispering Gallery Mode Sapphire resonators cooled into liquid helium dewars \cite {chang00,marra07,watabee06-eftf,ell04_lowdrift}. A detailed review of the techniques used in the design and construction of these CSOs has been published recently \cite{locke08}. Although this technology has already proved its ability to get the frequency stability specifications, it is not really suited for the ESA application which requires a large autonomy of the frequency reference. The objective of a continuous operation at least during 1 year rejects the use of a liquid helium dewar which has to be refilled periodically. Some attempts have been realized few years ago by using single stage
Gifford-McMahon cryocoolers \cite{hao99,uffc03_rutile}. Unfortunately, the
lowest temperature provided by these cryocoolers did not permit to get entirely
benefit of the sapphire resonator which presents an optimal temperature around
6K. Later J. Dick and R. Wang experimented a compensated sapphire resonator at
10K cooled into a two-stage Gifford McMahon cryocooler which approached the ESA
performances but still limited at $1\times 10^{-14}$ at $1$s and by a relatively
large drift at long term, {\it{i.e.}} $1\times 10^{-13}$/day \cite{wang99}. More
recently, Watabe {\it{et al}} proposed the use of a two-stage pulse tube cryocooler which presents a higher durability and a lower level of vibrations than other types of cryocooler but the performances they obtained still remain limited above $1\times 10^{-14}$ \cite{watabe03}.\\

This article presents the design, the breadboarding and the validation of a CSO which we named ``ELISA'', built under contract of the ESA by the Femto-ST Institute (F) as the prime contractor, the National Physical Laboratory (UK) and the TimeTech company  (D).

\clearpage
\section{ELISA sub-systems}

\begin{figure}[ht!]
	\centering
\scalebox{0.41}{\input{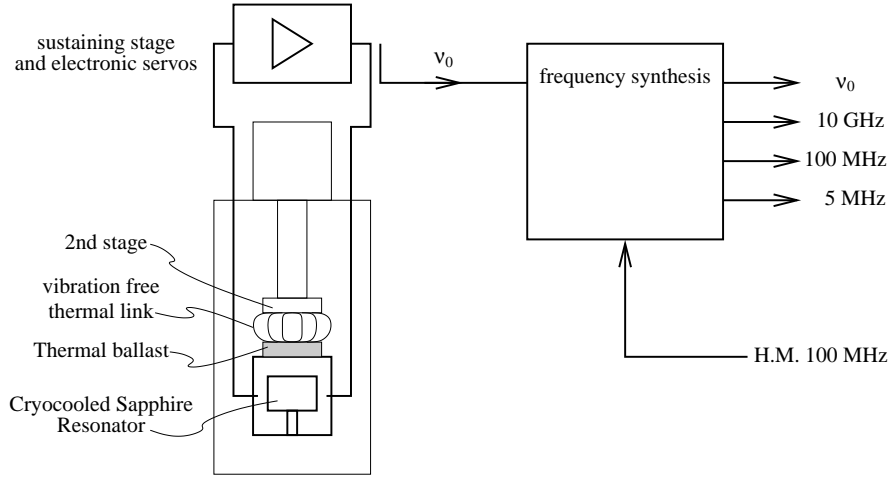}}
	\caption{\it{ELISA sub-systems. The frequency reference is a sapphire resonator maintained at a low temperature ($\approx 6$K) into a closed cycle cryocooler. The CSO delivers a signal at the frequency $\nu_{0}$ } which serves as reference for a frequency synthesis subsystem delivering the useful frequencies}
	\label{fig:rsi09-elisa-fig1}
	\end{figure}

The scheme in figure \ref{fig:rsi09-elisa-fig1} describes the main sub-systems constituting the Elisa frequency reference. 

The heart of the system is a whispering gallery mode sapphire resonator made of
a large and thick high purity sapphire (HEMEX grade \cite{www.crystalsystem}) cylinder placed in the center of a copper cavity (see section \ref{sec:reso}). This assembly is thermally connected to the second stage of a pulse tube cryocooler. A special \it{soft}\rm\ thermal link  and a thermal ballast (see section \ref{sec:cryo}) were designed in order to filter the vibrations  and the temperature modulation at about 1 Hz induced by the gas flow in the cryocooler.
 
Low thermal conductance coaxial cables are used to connect the Sapphire
resonator to the sustaining loop placed at room temperature. The oscillating
circuit is completed by two servos stabilizing the phase along the loop and the power injected into the resonator. These control loops use the same principle than those described in the reference \cite{locke08} (see section \ref{sec:osc}). %
Apart from the CSO signal at a frequency $\nu_{0}$, we have to generate three
other frequencies required for ESA applications: 10 GHz, 100 MHz and 5 MHz.
Moreover these useful signals should be phase locked at long term on a 100 MHz
reference coming from a Hydrogen Maser (HM). A frequency synthesis was then
designed to transfer the CSO's frequency stability to these useful signals with
a slight degradation at 5 MHz as a result of the performance of typical RF components. %
 In this paper, we will focus on the CSO only, the frequency synthesis, still not validated, will be detailed in a forthcoming paper.

\section{Sapphire resonator}
\label{sec:reso}

The frequency reference is a cylindrical sapphire resonator in which high order
modes called whispering gallery (WG) modes can be excited. These modes are
characterized by a high energy confinement in the dielectric due the total
reflexion at the vacuum-dielectric interface. As sapphire shows the lowest
dielectric losses in the microwave range, a Q-factor as high as 1 billion can be
obtained at the liquid-He temperature. The useful modes
are divided in two families: quasi-TM (WGH) and quasi-TE (WGE) modes.
They are further denoted by three integers: $m$, $n$ and $l$. $m$ is the number of wavelength in the azimuthal direction $\varphi$. $n$ and $l$ are the number of
field nodes in the radial and axial direction respectively. The modes of interest
have generally $10<m<20$ and $n=l=0$.  In pure sapphire mono-crystal, WG modes
will have a monotonic frequency-vs-temperature law. 
The pure sapphire resonator has not th efrequency-vs-temperature turning point otherwise found in most piezoelectric resonators after appropriate design.  Although the mode
frequency thermal sensitivity decreases significantly at low temperature, it
never goes low enough for the target stability to be achieved with state-of-the-art temperature control.
Fortunately, it turns out that high-purity sapphire crystals always contain a small
concentration of paramagnetic impurities, as Cr$^{3+}$, Fe$^{3+}$ or Mo$^{3+}$.
These ions induce a small magnetic permeability whose temperature dependence
compensates at a given temperature  $T_{0}$ for the natural sapphire resonator
thermal sensitivity. This turnover temperature $T_{0}$ depends on the mode and
on the impurity concentration but for many of the resonators tested by different
groups, the turnover temperature is generally in the range $5-8$K. This thermal compensation relaxes dramatically the thermal stabilization requirements. Indeed, with such a thermal compensation phenomena, the frequency stability objective of $3\times 10^{-15}$ can be obtained with a $\pm 1$ mK thermal stabilization which can be easily fulfilled with commercial cryogenic temperature controllers.\\

\begin{figure}[ht!]
	\centering
	\includegraphics[scale=0.85]{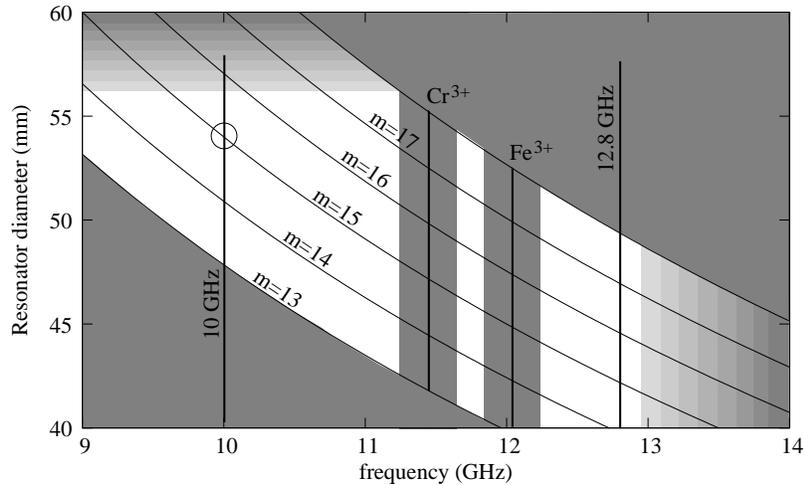}
	\caption{\it{Resonator frequency selection rules. The chosen resonator parameters correspond to the point highlighted by the circle: frequency $\nu_{0}=10$ GHz, diameter $D \approx 54$mm, azimuthal number $m=15$}}
	\label{fig:choix-frequency}
	\end{figure}

The resonator size and WG mode order determine the resonant frequency and the unloaded Q-factor. 
Experience shows that the best results are obtained with a resonator presenting
a ratio diameter over height of the order of $D/H \approx 3/5$ \cite{Krupka96} and with WG modes between 13 and 18.
The energy confinement in the dielectric improves as the mode order increases.  This fact has two practical consequences: 1) $Q$ is progressively degraded at lower-order modes (less than 13) because of electromagnetic radiation, and 2) high order modes (greater than 18) are difficult or impossible to exploit because the couplers need too sharp adjustment.   
Another difficulty connected with high-order modes is the presence of many spurious modes, which makes the frequency selection difficult.
For practical reasons the resonator size can not be too large.  A resonator diameter $D$ of about 50 mm is comfortable for mechanics and cryogenics. This means that the resonant frequency should not be lower than some 9 GHz.
Dielectric dissipation in sapphire increases as frequency increases.  This phenomenon sets a soft upper limit at some 13 GHz.  This limit is about independent of the resonator size.
Finally, the resonance of the Cr$^{3+}$ ion at 11.45 GHz, and the resonance of the Fe$^{3+}$ ion and 12.04 GHz are to be avoided.  A margin of $\pm200$ MHz is recommended.
Figure~\ref{fig:choix-frequency} summarizes the frequency selection rules we adopted. This figure represents the relation between the resonator diameter and its resonance frequency for the WGH mode family assuming $D/H=3/5$. 

\subsection{Design strategy}\label{ssec:design-strat}

It is generally agreed that the resonator is the core of the oscillator, which determines the frequency stability.  Provided the criteria discussed at the beginning of Section~\ref{sec:reso} are met, the most critical parameters are the quality factor $Q$ and the thermal stability of the natural frequency in the vicinity of the turnover temperature. 
As a consequence, the physicist is inclined to start from a set of available resonators, identify the highest-$Q$ mode, check on the thermal stability, and build the entire oscillator around this empirical choice.
One problem with this approach is that the frequency synthesizer, needed for the oscillator to deliver a suitable round frequency, has to provide an interpolation frequency in a wide range, up to 1 GHz.  Implementing such synthesizer with micro-Hertz resolution, and with stability and spectral purity high enough not to degrade the output signal, is not simple.  Another problem is that each resonator has its own optimal frequency, which determines the design of the electronics.  This turns into a technical difficulty if more than one oscillator will be needed in the future, or if the resonator has to be replaced for any reason.

Aware of all these problem, we opted for a different approach based on the criteria discussed underneath.
After refining the electromagnetic model \cite{esa-workshop08}, we are able to design a resonator with a high-$Q$ mode at the frequency $\nu_0$ of our choice.  The electrical and mechanical tolerances yield a frequency accuracy of $5{\times}10^{-4}$ ($\pm5$ MHz at 10 GHz) and a reproducibility of $10^{-4}$ ($\pm1$ MHz at 10 GHz).
That said, we chose a resonant frequency $\nu_0=\nu_{00}-\delta_{RF}$, where $\nu_{00}$ is a comfortable round value, and $\delta_{RF}=10\pm5$ MHz.  This choice is justified by the fact that a single-chip direct digital synthesizer (DDS) has the desired resolution and provides sufficient stability and spectral purity if the output frequency does not exceed some 15--20 MHz.  Additionally, $\delta_{RF}$ cannot be too low frequency ($<5$ MHz), otherwise it is difficult to avoid the spur of frequency $\nu_0$, too close to $\nu_{00}$.
In this conditions, a DDS fits the needs for the correction $\delta_{RF}$ 
The nominal size of the resonator makes $\nu_0<\nu_{00}$, so the correction $-\delta_{RF}$ takes a negative value.   It is still possible to thin the resonator so that $\nu_0=\nu_{00}+\delta_{RF}$, which can fix a number of mistakes without modifying the electronics. 

Figure~\ref{fig:choix-frequency} suggests that $\nu_{00}$ can be either 10 GHz or 12.8 GHz, so that a low-noise 100 MHz output can be obtained from a cascade of by-10 or power-of-2 frequency dividers.
The value $\nu_{00}=10$ GHz has been chosen because low-flicker SiGe amplifiers \cite{aml,rubiola-phase-noise} are available at this frequency, while 12.8 GHz falls beyond a sharp cutoff.  

Our design strategy has three additional merits connected to the low value of the interpolation frequency $\delta_{RF}$.  The first one is that it simplifies the synthesizer.  The second one is that the synthesizer can be over-engineered for high stability and low noise at a reasonably low cost.  If a lucky oscillator exceeds the expected stability, the benefit is not lost.    
The third one is that the synthesizer can be tested without the oscillator, which simplifies the logistics and speeds up the validation process.

\subsection{Resonator realization and validation}

\begin{figure}[h]
\centering
\includegraphics[width=0.65\linewidth]{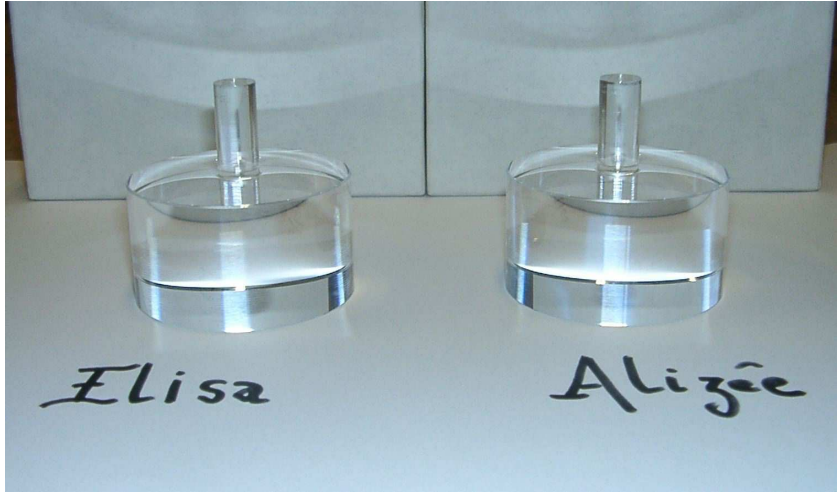}
 \caption{ \it{The two HEMEX grade sapphire resonators.}}
 \label{fig:vg-elisa-alizee-picture}
\end{figure}

The key parameters to accurately determine the modes frequencies are the values of the sapphire tensor permittivity components used for the calculation. By comparing experimental frequencies of available crystals and simulation results, we deduced the values of the permittivity components at 4K to be: $\epsilon_{\perp}=9.270,688$ and $\epsilon_{/\!/}= 11.340,286$.
Eventually, by using a  Finite Elements analysis, we get a $WGH_{15,0,0}$ mode
at 9.99 GHz with $D= 54.2$ mm and \mbox{$H=30$ mm}. Taking into account the machining possibilities and the cost, we eventually ordered two sapphire pieces with $
 D = 54.2\mathrm{mm} \pm 10\mu\mathrm{m~~and~~}  H = 30\mathrm{mm} \pm 20\mu\mathrm{m}$.
The resonator frequency is then defined as $9.99$ GHz $\pm3.5$ MHz.\\
The resonator has a spindle (diameter 10 mm, length 22 mm) to be clamped from below. Such a mounting enables to reduce the mechanical stress on the region where the electromagnetic field is confined \cite{chang01-ifcs}. Two quasi identical resonators were machined from the same HEMEX sapphire boule by the Crystal System company. These two resonators named Elisa and Aliz\'ee are shown on the figure \ref{fig:vg-elisa-alizee-picture}.\\

\begin{figure}[ht!]
	\centering
	\includegraphics[scale=0.4]{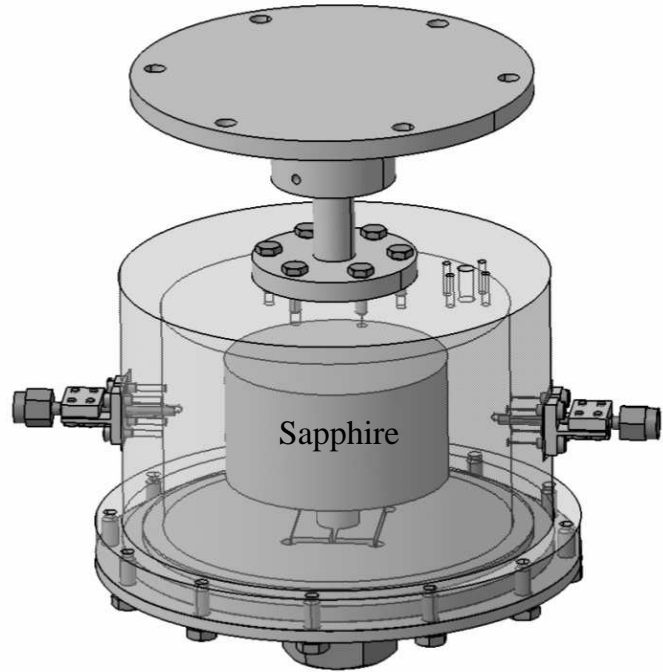}
	\caption{\it{ELISA resonator design.}}
	\label{fig:elisa-cavity}
	\end{figure}

The cryogenic microwave resonator making up the Elisa's frequency reference is represented  in the figure \ref{fig:elisa-cavity}.\\

The resonator is clamped on the inferior plate closing the gold plated copper
cavity. The latter is maintained in thermal contact with the cooling source through a copper piece in which a thermal sensor and a heater are anchored. 
To couple the resonator to the external circuit, we use two small magnetic loops intercepting the $H_{\varphi}$ magnetic field component of the resonator. By varying the penetration depth of the loop into the cavity, we can adjust the input ($\beta_{1}$) and output ($\beta_{2}$) coupling coefficients. To maximize the sensitivity of the frequency discriminator used to control the phase along the oscillating loop, $\beta_{1}$ has to be set near the unity. On the contrary $\beta_{2}$ has to be set to a low value to not degrade the loaded Q-factor, but sufficiently high to limit the resonator insertion losses ($IL$). Table \ref{tab:resonator-parameters} summarizes the main characteristics of the two resonators at their turnover temperature $T_{0}$. The adjustment of the coupling coefficients have been obtained after few cool-downs. 

\begin{table}[h!!!!!]
\centering
\caption{Resonators parameters}
\label{tab:resonator-parameters}
\begin{tabular}{lcccccc}
 \hline 
& ~~~$\nu_{0}$ (GHz)~~~ &~~~ $T_{0}$ (K)	 ~~~& ~~~$Q_{L}$ ~~~ & ~~~$\beta_{1}$ ~~~& ~~~$\beta_{2}$~~~&~~~ $IL$ (dB)~~~\\

  \hline 
ELISA 	& $9.989,121$ 	&6.13 	& $7.4\times 10^{8}$ 	& 1 	& $\approx 0.02$ & $-24.9$\\
 \hline 
ALIZEE 	& $9.988,370$ 	& 6.05		& $6.9 \times 10^{8}$ 	& 1.09 	&
$\approx 0.02$ & -26 \\
 \hline 
\end{tabular}
\end{table}

\begin{figure}[ht!]
\centering
\scalebox{0.76}{\input{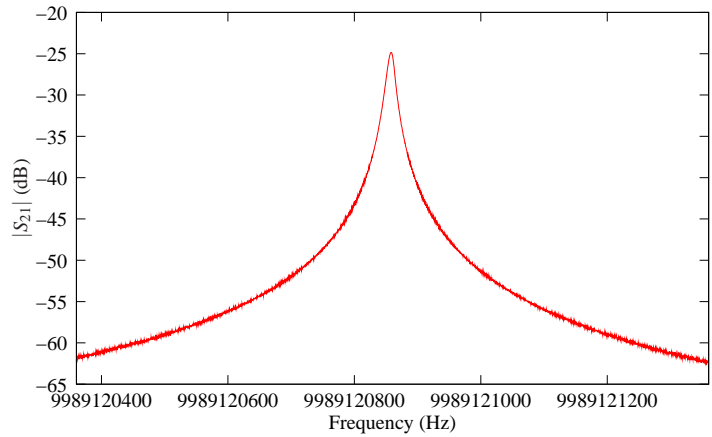}}
\caption{\it{ELISA transmission coefficient around the $WGH_{15,0,0}$ mode frequency. Frequency span 1 kHz.}}
\label{fig:grop-rsi09-elisa-s21}
\end{figure}

The figure \ref{fig:grop-rsi09-elisa-s21} shows the modulus of the Elisa transmission coefficient around the $WGH_{15,0,0}$ mode frequency in a 1 kHz span. The resonator bandwidth is 14 Hz.

\section{CRYOCOOLER}
\label{sec:cryo}

To achieve the required oscillator frequency stability, the oscillator's sapphire resonator must be maintained in a cryogenic environment that is sufficiently free of mechanical vibration and where the resonator's temperature is uniform and precisely controlled at a specific value. Two options were available at the beginning of the ELISA project:  using a liquid helium cryostat or a cryogen-free cryocooler-based cryostat.\\

With a bath cryostat, the vacuum can in which the resonator resides is
surrounded by a bath of cryogenic liquid (in this case liquid helium); within a
dewar. Such an arrangement naturally provides the sapphire resonator with a
thermally uniform environment, relatively free of mechanical vibration (except
from bubbling). Cryostats based on cryocoolers, on the other hand, will
generally subject the resonator to unacceptable levels of mechanical vibration
and thermal non-uniformity ({\it{i.e.}} temperature gradients) unless the cryostat is designed in such a way that these perturbations are suppressed.\\

The key point that made us opt for the cryocooler option is that unlike bath cryostat, they don't need regular refills of liquid helium. These refillings disturb the sapphire resonator which needs around two days to recover. Nevertheless two problems had to be solved: the vibrations level and the temperature fluctuations of the cryocooler cold stage. We measured  temperature fluctuations of $\pm 100$ mK on a Cryomech PT405 with a 1.3 Hz frequency corresponding to the gas cycle in the system.\\

\begin{figure}[ht!]
\centering
\includegraphics[scale=0.36]{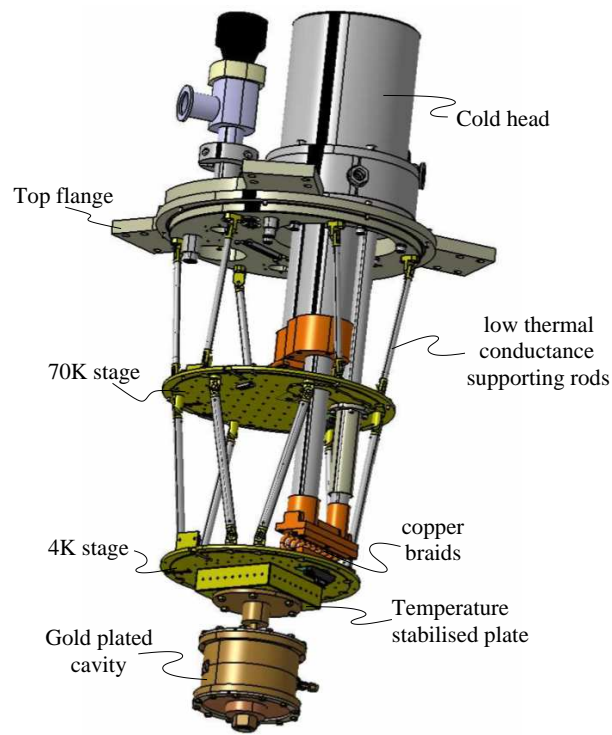}
\caption{\it{Crycooler internal design.}}
\label{fig:nb-cryo-internal}
\end{figure}

Oxford Instruments supplied a 4K Cryofree$^{\circledR}$ cryostat modified to achieve the performance required by the ELISA project. Those were a low displacement on the experiment plate (less than two microns in three axes) with a temperature stabilization of $\pm1$ mK over 1000 s and a cooling power at 4K of 50 mW. The cryocooler used by Oxford Instruments is a two-stage pulse-tube refrigerator with a rotary valve decoupled from the main dewar, eliminating the need for moving parts in the cold head. The cryostat design of the vibration reduction system is similar to those described in the references \cite{tomaru05,caparelli06}. The Oxford Instruments design is represented in figure \ref{fig:nb-cryo-internal}. \\

\begin{figure}[ht!]
\centering
\includegraphics[scale=0.85]{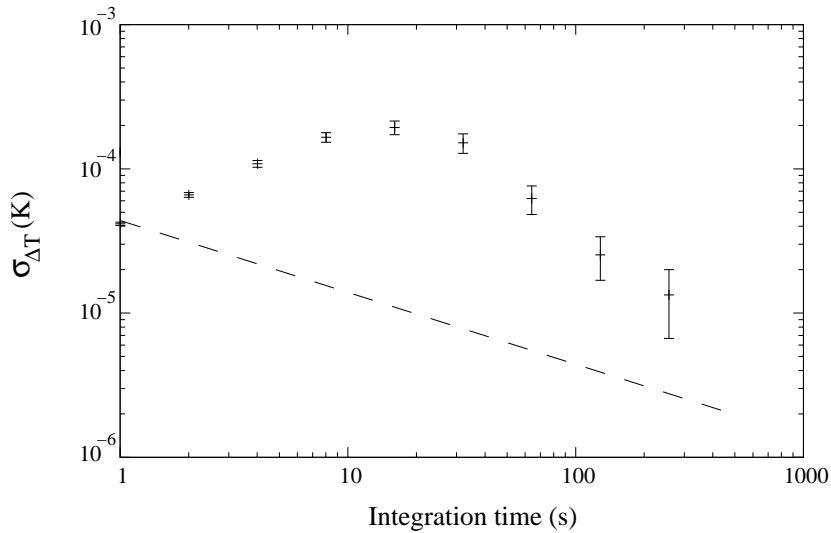}
\caption{\it{Measured temperature stability on the experimental plate. The dashed line represents the temperature measurement resolution.}}
\label{fig:oi-temperature-ADEV}
\end{figure}

The two stages of the cryostat (77K shield and 4K cold plate) are thermally linked to the cryocooler stages with floppy copper heat links. The support thin-wall tubes are mounted like an hexapod to give rigidity to the system. To limit the temperature fluctuations of the cold stage, a gadolinium gallium garnet (GGG) crystal \cite{dal88} was mounted between this stage and a copper temperature stabilization block supporting the experiment using four legs. The four legs provided a tuned weak thermal link to again optimize the temperature stability and cooling power
of the cryostat. Few combinations of materials for the legs were used and the optimized performance was obtained when using two stainless steel and two brass/copper legs. To check the temperature stability on the experimental plate, we fixed on it an independent high resolution thermal sensor and recorded the measured temperature. Calculating the Allan deviation of the recorded samples we get the temperature versus integration time stability showed in figure \ref{fig:oi-temperature-ADEV}. \\

Although we observed  a bump on the temperature stability curve around an
integration time of about 20 s, the temperature stability requirement,
{\it{i.e}} $\Delta T \leq 1$ mK rms, is totally fulfilled.  \\

A three axis accelerometer was  fixed on the experimental plate and we measured less than 1 $\mu$m vibration in $x,~y$ and $z$ direction \cite{eftf2009-elisa}. The base temperature at stabilization stage with the customer wiring is 6K with 50 mW applied to the stabilization stage.

\section{OSCILLATOR}
\label{sec:osc}

\begin{figure}[ht!]
\centering
\scalebox{0.35}{\input{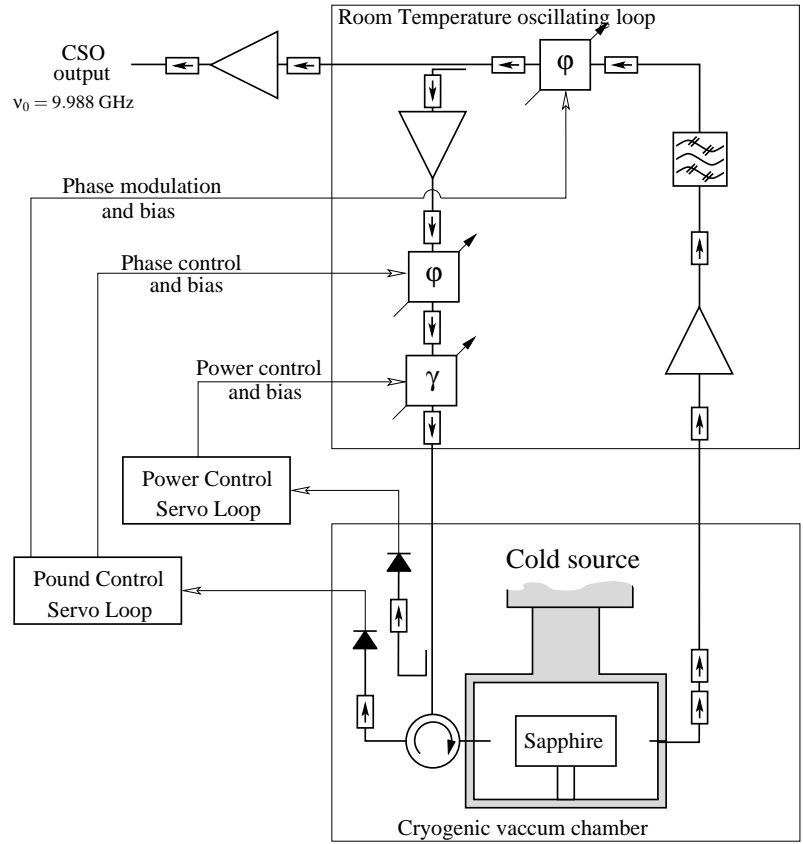}}
\caption{\it{Elisa oscillator design. The sustaining loop is completed with two additional servo loops stabilizing the phase of the circulating signal and the power injected inside the resonator.}}
\label{fig:osc-loop}
\end{figure}

The oscillator circuit is described in the figure \ref{fig:osc-loop}. It is a classical transmission oscillator circuit with two additional servo loops that control the phase and the power of the circulating signal. These two servos are mandatory to get a high frequency stability. 

Without these two servo loops, the CSO frequency stability will stay limited to some $1\times 10^{-13}$ at short term and will be furthermore degraded at long integration time due to the environmental sensitivity of the circuit. 
The first servo loop is based on the Pound frequency discriminator principle
\cite{drever83,galani84,black01}, it ensures that the CSO oscillates at the resonator
frequency $\nu_{0}$ by compensating any variation of the phase lag along the
loop. It uses a phase modulation at a frequency of the order of few tens of kHz
to probe the resonance. The phase modulation is applied to the microwave signal
through a a first Voltage Controlled Phase Shifter (VCPS 1). A lock-in amplifier
demodulates the signal reflected by the resonator to generate an error signal
which is eventually added to the dc-bias of a second voltage controlled phase
shifter (VCPS 2). Due to the radiation pressure and to the self resonator
heating,  the resonator frequency presents a power sensitivity of the order of
$4\times 10^{-11}/$mW. The power servo loop ensures that the power injected into
the resonator stays constant. A tunnel diode placed as near as possible to the
resonator input enables to get a voltage proportional to the signal power. This
voltage
is compared to a high stability voltage reference and the resulting error signal is used to control the bias of a voltage controlled attenuator. As explained in \cite{locke08} the modulating signal applied to VCPS 1 induces as well an undesirable  amplitude modulation (AM) which shifts the oscillating frequency from $\nu_{0}$. As the frequency shift is dependant on the AM index, the CSO frequency stability will be limited by the variations of the VCPS insertion losses which in turn couples the CSO frequency to environment and especially to the room temperature. A solution has been proposed to surpass this last limitation \cite{luiten-PhD,locke08}, but at this stage of development we did not yet implement such a supplementary control on our CSO. The result we obtain demonstrate that this effect did not affect the CSO frequency stability above $3 \times 10^{-15}$.

\section{FREQUENCY PERFORMANCES}
\label{sec: perfo}

To evaluate the relative frequency stability of Elisa, it has been compared with
a second CSO build around the second resonator Aliz\'ee which is cooled in a
liquid helium dewar. Apart from the cooling method, the two CSOs are quasi identical.

\begin{figure}[ht!]
\centering
\scalebox{0.9}{\input{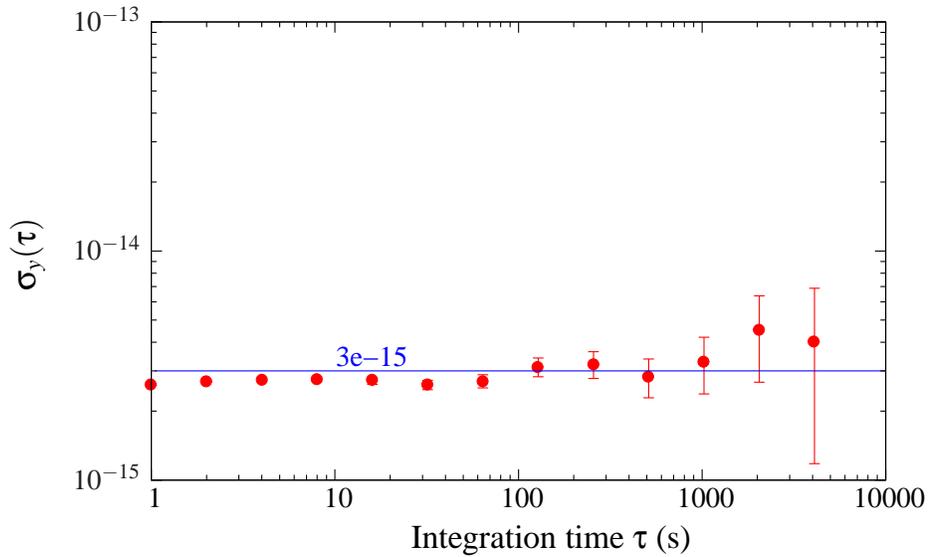}}
\caption{\it{ELISA frequency maximal frequency instability measured by direct comparison with a  liquid helium cooled sapphire oscillator.}}
\label{fig:grop-rsi09-sigma_y}
\end{figure}

Before running Elisa into the cryocooler, some preliminary measurements have been realized with the two resonators cooled in liquid helium dewar \cite{eftf2009-elisa}.  The obtained results have demonstrated
that the liquid helium cooled CSO presents a short term relative frequency instability less or equal to $3\times 10^{-15}$. Any degradation in the Elisa's frequency stability due to some residual mechanical vibrations should then be detected.
The measurement technique is a standard one: the two oscillator signals are
mixed to get a beatnote at the frequency difference which is of about 750 kHz. The
beat signal is directly send to a high resolution counter and the Allan
deviation $\sigma_{y}$ is computed from data averaged over 1s. The figure
\ref{fig:grop-rsi09-sigma_y} shows $\sigma_{y}$ as a function of the integration
time $\tau$. The $3\times 10^{-15}$ frequency stability specification is almost
met for $\tau \leq 1,000$ s. This result corresponds to a data recording of about 7 hours without any post-processing applied to the raw data.

A rapid view on this Allan deviation curve shows that the measured frequency stability appears limited at short term by a ``plateau`` just below $3\times 10^{-15}$. We are confident that the observed plateau does not correspond to a flicker frequency noise but more certainly to the superposition of some residual frequency deterministic variations. We detected recently that the thermostat stabilizing the voltage references used both for power and frequency servo suffer from a small but detectable pumping at a rate of the order of 300s. Power servos which are more sensible to the voltage reference noise have been modified but the same issue still subsists in the Pound loops. That can explain the little bumps in the $\sigma_{y}$ curve around  300 s.\\

\begin{figure}[ht!]
\centering
\includegraphics[scale=0.86]{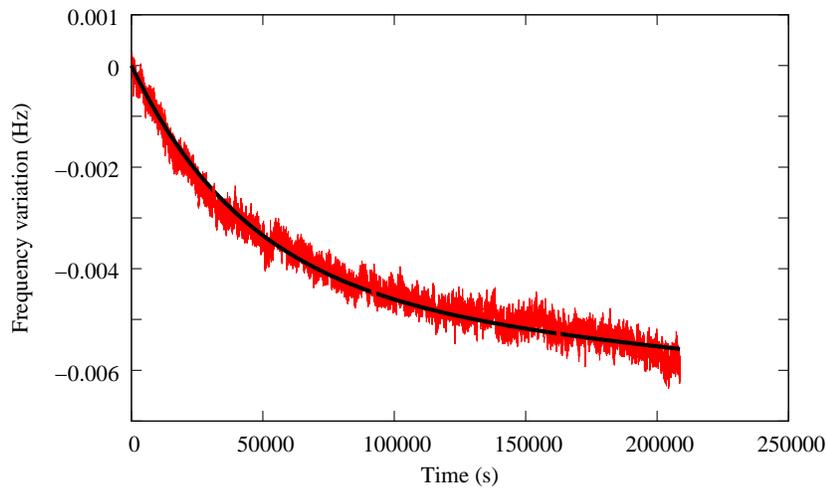}
\caption{\it{Evolution of the beat-note frequency after refilling.}}
\label{fig:grop-rsi09-beatnote}
\end{figure}

When two almost equivalent oscillators are directly compared, it is a common practice to divide the measured Allan deviation by $\sqrt{2}$ assuming the two oscillator noises are equivalent and uncorrelated. We did not applied this rule in our case because, although build with practically identical resonators, the two CSOs behave differently. All the parameters related to the servo loops as the integrator outputs and the lock-in output noise, are much more stable for the crycooled CSO. 
In the case of the liquid helium dewar CSO, the sustaining loop has a length of about 4 m. The cables linking the resonator to the sustaining circuit pass through the helium bath. Due to the helium evaporation their electrical length and loss are continuously varying.  Moreover, the almost critical input coupling that we fortunately get for Elisa should limit the effect of AM-index fluctuations \cite{locke08}. Eventually the liquid helium dewar has to be refilled every week. This operation pertubes the CSO frequency which takes time to stabilize. As helium evaporates continuously we doubt that the liquid helium CSO get its equilibrium between two refillings. The figure \ref{fig:grop-rsi09-beatnote} shows the beatnote frequency variations after refilling.\\

We note that more than one day is needed to get the best estimation of the frequency stability. Nevertheless, after 200,000 s the residual frequency drift is less than $5\times 10^{-14}/$ day.\\

Eventually although there are obviously still some margin to improve the result,
it has been proved that the crycooled CSO can meet the frequency stability
specification of $3\times 10^{-15}$ up to 1,000 s integration times.



\begin{thebibliography}{10}

\bibitem{chang00}
S.~Chang, A.~G. Mann, and A.~N. Luiten, ``Improved cryogenic sapphire
  oscillator with exceptionally high frequency stability,'' {\em Electronics
  Letters}, vol.~36, pp.~480--481, Mar.~2~ 2000.

\bibitem{marra07}
G.~Marra, D.~Henderson, and M.~Oxborrow, ``Frequency stability and phase noise
  of a pair of x-band cryogenic sapphire oscillators,'' {\em Meas. Sci.
  Technol.}, vol.~18, pp.~1224--1228, 2007.

\bibitem{watabee06-eftf}
K.~Watabe, J.~Hartnett, C.~R. Locke, G.~Santarelli, S.~Yanagimachi, T.~Ikegami,
  and S.~Ohshima1, ``Progress in the development of cryogenic sapphire
  resonator oscillator at nmij/aist,'' in {\em Proc.\ 20th European Frequency
  and Time Forum}, (Braunschweig, Germany), pp.~92--95, march. 27--30 2006.

\bibitem{ell04_lowdrift}
P.~Y. Bourgeois, F.~Ladret-Vieudrin, Y.~Kersal\'e, N.~Bazin, M.~Chaubet, and
  V.~Giordano, ``Ultra low drift microwave cryogenic oscillator,'' {\em
  Electronics Letters}, vol.~40, pp.~605--606, May~13~ 2004.

\bibitem{locke08}
C.~R. Locke, E.~N. Ivanov, J.~G. Hartnett, P.~L. Stanwix, and M.~E. Tobar,
  ``Design techniques and noise properties of ultrastable cryogenically cooled
  sapphire-dielectric resonator oscillators,'' {\em Review of Scientific
  Instruments}, vol.~79, pp.~051301--1--12, 2008.

\bibitem{hao99}
L.~Hao, N.~Klein, J.~C. Gallop, W.~J. Radcliffe, and I.~S. Ghosh, ``Temperature
  compensated cryogenic whispering gallery mode resonator for microwave
  frequency standard applications,'' {\em IEEE Transactions on Instrumentation
  and Measurement}, vol.~48, pp.~524--527, april 1999.

\bibitem{uffc03_rutile}
Y.~Kersal\'e, S.~Vives, C.~Meunier, and V.~Giordano, ``Cryogenic monolithic
  sapphire-rutile temperature compensated resonator oscillator,'' {\em IEEE
  Transactions on Ultrasonics, Ferroelectrics and Frequency Control}, vol.~50,
  pp.~1662--1666, Dec. 2003.

\bibitem{wang99}
R.~T. Wang and G.~J. Dick, ``Cryocooled sapphire oscillator with ultrahigh
  stability,'' {\em IEEE Transactions on Instrumentation and Measurement},
  vol.~48, pp.~528--531, april 1999.

\bibitem{watabe03}
K.~Watabe, Y.~Koga, S.~Ohshima, T.~Ikegami, and J.~Hartnett, ``Cryogenic
  whispering gallery sapphire oscillator using {4K} pulse-tube cryocooler,'' in
  {\em Proc. of the 2003 IEEE International Frequency Control Symposium},
  (Tampa, Fl., USA), pp.~388--390, IEEE, New York, 1992, May~4--8 2003.

\bibitem{www.crystalsystem}
\texttt{http://www.crystalsystem/}.

\bibitem{Krupka96}
J.~Krupka, D.~Cros, A.~Luiten, and M.~Tobar, ``Design of very high q sapphire
  resonators,'' {\em Electronics Letters}, vol.~32, pp.~670--671, 1996.

\bibitem{esa-workshop08}
S.~Grop, V.~Giordano, P.~Bourgeois, Y.~Kersal\'e, N.~Bazin, M.~Oxborrow,
  G.~Marra, C.~Langham, E.~Rubiola, and J.~{De Vicente}, ``{ELISA:} an
  ultra-stable frequency reference for space mission ground segment,'' in {\em
  Microwave Technology and Techniques Workshop 2008 - Innovations and
  Challenges.}, ({ESA/ESTEC Noordwijk, The Nederlands}), p.~06\_Grop, 6-7 May
  2008.

\bibitem{aml}
\texttt{http://www.amlj.com/lowphasenoise.html/}.

\bibitem{rubiola-phase-noise}
E.~Rubiola, {\em Phase noise and frequency stability in oscillators}.
\newblock Cambridge University Press, 2008.
\newblock ISBN 978-0-521-88677-2.

\bibitem{chang01-ifcs}
S.~Chang and A.~Mann, ``Mechanical stress caused frequency drift in cryogenic
  sapphire resonator,'' in {\em Proc.\ of the 2001 IEEE International Frequency
  Control Symposium}, (Seattle, WA, USA), pp.~710--714, june 6-8 2001.

\bibitem{tomaru05}
T.~Tomaru, T.~Suzuki, T.~Haruyama, T.~Shintomi, N.~Sato, A.~Yamamoto,
  Y.~Ikushima, R.~Li, T.~Akutsu, T.~Uchiyama, and S.~Miyoki, {\em Cryocoolers
  13}, ch.~Vibration-Free Pulse Tube Cryocooler System for Gravitational Wave
  Detectors, Part I: Vibration-Reduction Method and Measurement, pp.~695--702.
\newblock Springer {US}, 2005.

\bibitem{caparelli06}
S.~Caparelli, E.~Majorana, V.~Moscatelli, E.~Pascucci, M.~Perciballi, P.~Puppo,
  P.~Rapagnani, and F.~Ricci, ``Vibration-free cryostat for low-noise
  applications of a pulse tube cryocooler,'' {\em Review of Scientific
  Instruments}, vol.~77, pp.~095102--1--7, 2006.

\bibitem{dal88}
W.~Dal, E.~Gmelin, and R.~Kremer, ``Magnetothermal properties of sintered
  {G}d$_{3}${G}a$_{5}${O}$_{12}$,'' {\em J. Phys. D: Appl. Phys}, vol.~21,
  pp.~628--635, 1988.

\bibitem{eftf2009-elisa}
S.~Grop, P.~Bourgeois, N.~Bazin, C.~Langham, M.~Oxborrow, J.~D. V.~E. Rubiola,
  Y.~Kersal\'e, and V.~Giordano, ``{ELISA} : An ultra-stable oscillator for esa
  deep space antennas,'' in {\em Proc.\ of the joint meeting IFCS-EFTF},
  (Besan\c con, France), pp.~376--380, april 20-24 2009.

\bibitem{drever83}
R.~Drever, J.~Hall, F.~Kowalski, J.~Hough, G.~Ford, A.~Munley, and H.~Ward,
  ``Laser phase and frequency stabilization using an optical resonator,'' {\em
  Appl. Phys. B}, vol.~31, pp.~97--105, 1983.

\bibitem{galani84}
Z.~Galani, M.~Bianchini, W.~Jr., C.~Raymond, R.~Dibiase, R.~Laton, and
  J.~{Bradford Cole}, ``Analysis and design of a single-resonator gaas fet
  oscillator with noise gegeneration,'' {\em IEEE Transactions on Microwave
  Theory and Techniques}, vol.~32, pp.~1556--1565, december 1984.

\bibitem{black01}
E.~Black, ``An introduction to {P}ound-{D}rever-{H}all laser frequency
  stabilization,'' {\em Am. J. Phys}, vol.~1, pp.~79--87, Jan. 2001.

\bibitem{luiten-PhD}
A.~Luiten, {\em Sapphire secondary frequency standards}.
\newblock PhD thesis, University of Western Australia, Nedlands, 1995.

\end{thebibliography}
\end{document}